\documentclass[a4paper,11pt]{article}
\pdfoutput=1 

\usepackage{jcappub,color} 
\usepackage{amsmath,amsfonts,graphics,epsfig,bm}
\usepackage[T1]{fontenc} 
\newcommand{\be}{
\[
}
\newcommand{\ee}{
\]}

\newcommand{\blue}{\color{blue}}
\newcommand{\black}{\color{black}}
\renewcommand{\blue}{\black}

\title{Constraining the photon coupling of ultra-light dark-matter
axion-like particles
by polarization variations of parsec-scale jets in active galaxies}

\author[a,b,c]{M.M.~Ivanov,}
\author[d,e]{Y.Y.~Kovalev,}
\author[f]{M.L.~Lister,}
\author[a,e]{A.G.~Panin,}
\author[g,d]{A.B.~Pushkarev,}
\author[h,i]{T.~Savolainen,}
\author[a,1]{and S.V.~Troitsky\note{Corresponding author.}}
\affiliation[a]{
\footnotesize Institute for Nuclear
Research of the Russian Academy of Sciences,
60th October Anniversary
Prospect 7a, Moscow 117312, Russia}
\affiliation[b]{
\footnotesize School of Natural Sciences, Institute for Advanced Study,
1 Einstein Drive, Princeton, NJ 08540,  USA
}
\affiliation[c]{
\footnotesize
Institute of Physics, Laboratory of Particle Physics and Cosmology (LPPC),
\'Ecole Polytechnique F\'ed\'erale de Lausanne, CH-1015, Lausanne,
Switzerland }
\affiliation[d]{
\footnotesize Astro Space Center, P. N. Lebedev Physical Institute,
Russian Academy of Sciences, Profsoyuznaya 84/32, Moscow, 117997, Russia
}
\affiliation[e]{
\footnotesize
Moscow Institute of Physics and Technology,
Institutsky per.\ 9, Dolgoprudny, 141700, Russia
}
\affiliation[f]{
\footnotesize\blue Department of Physics and Astronomy, Purdue University,
525 Northwestern Avenue, West Lafayette, IN 47907, USA }
\affiliation[g]{
\footnotesize
Crimean Astrophysical Observatory, Russian Academy of Sciences, Nauchny,
298409, Russia
}
\affiliation[h]{
\footnotesize\blue  Aalto University Department of Electronics and
Nanoengineering, PL 15500, 00076 Aalto, Finland}
\affiliation[i]{
\footnotesize\blue Aalto University Mets\"ahovi Radio Observatory,
Mets\"ahovintie 114, 02540 Kylm\"al\"a, Finland
}

\black

\emailAdd{st@ms2.inr.ac.ru}

\date{November 27, 2018}

\abstract{%
Ultra-light dark matter may consist of axion-like particles with masses
below $10^{-19}$~eV. Two-photon interactions of these particles affect the
polarization of radiation propagating through the dark matter. Coherent
oscillations of the Bose condensate of the particles induce periodic
changes in the plane of polarisation of emission passing through the
condensate.
\blue
We estimate this effect and analyze MOJAVE VLBA polarization observations
of bright downstream features in the parsec-scale jets of active galaxies.
Through the non-observation of periodic polarization changes, we are able
to constrain \black the photon coupling of the ultra-light dark-matter axion-like
particles at the level of
$\lesssim 10^{-12}$~GeV$^{-1}$ for masses between $\sim 5\times
10^{-23}$~eV and $\sim 1.2 \times 10^{-21}$~eV.
}

\begin{document}
\maketitle
\flushbottom
\section{Introduction}
\label{sec:intro}
Despite impressive experimental efforts towards direct-detection, collider
and indirect searches for weakly interacting massive particles (WIMPs, see
e.g.\ Refs.~\cite{direct-review, collider-review, indirect-review} for
corresponding reviews), no confirmation of the existence of a candidate
dark-matter (DM) particle of this kind has been obtained. As a result, we
witness a growing interest to non-WIMP DM candidates, including axions,
axion-like particles, sterile neutrinos etc. On the other hand, the
conventional Cold Dark Matter (CDM) scenario suffers from considerable
tensions with observational data regarding structure formation at small
(below kiloparsec) scales, see e.g.\ Ref.~\cite{1306.0913} for a review.
These tensions include the ``missing satellites'' \cite{9901240, 9907411},
``too-big-to-fail'' \cite{1103.0007} and ``cusp-core'' \cite{0708.1492}
problems. In general, CDM, as a cold, scale-free and non-interacting
substance, forms too much structure at small scales.

One of the scenarios put forward to overcome these tensions is based on
the concept of ultra-light (UL), also called ``fuzzy'', dark matter (see
e.g.\ Refs.~\cite{002495, 0003018, 0003365}, but also early pioneering
works
\cite{Turner1983, Tkachev1986, Khlopov1985} and, for reviews, \cite{Doddy,
1610.08297}). In this approach, the DM particle is so light that its de
Broglie wavelength is of the order of a kiloparsec, the ``problematic''
scale of structure formation. This corresponds to a DM particle mass of
order of $m \sim 10^{-22}$~eV. With this low mass, the observed DM energy
density requires very high number densities of the ULDM particles, which
imply that they exist in the form of a classical bosonic field, or a Bose
condensate.  To protect the low mass from radiative corrections, it is
often assumed that the ULDM field corresponds to a pseudo-Goldstone boson
of some broken symmetry, just like the Quantum Chromodynamics
(QCD) axion is related to breaking of the Peccei-Quinn symmetry. Related to
QCD or not, this particle develops a non-renormalizable coupling to
photons similar to that of the axion. Pseudoscalars with these couplings
are called axion-like particles (ALPs, see e.g.\ Ref.~\cite{Ringw-review}
for a review). They appear in a natural way in many extensions of the
Standard Model of particle physics, including those related to string
theory \cite{Wit, Dub, Ringw}.

The photon coupling of ALPs makes it possible to search for their
manifestations in laboratory experiments and in astrophysical
environments, see e.g.\ Refs.\
\cite{Doddy, 1801.08127, 1708.02111, 1612.01864} for recent reviews. A
range
of approaches are based on the ALP-photon mixing
in external magnetic fields \cite{RaffeltStod}, which
results in the ALP-photon oscillations, the ALP Primakoff effect and
vacuum birefringence. Here, we follow a different approach which is based
on the same ALP-photon interaction: the polarization properties of light
are changed when it propagates in the external pseudoscalar field.

The condensate of ULDM particles is naturally produced in the early
Universe, as follows both from analytical estimates and from detailed
numerical simulations. The produced condensate forms domains of the size
of order $\sim 100$~pc (see e.g.\ Ref.\ \cite{Schive} for a numerical
demonstration). Within each clump, the ALP field experiences fast coherent
oscillations with the period determined by $m$. These oscillations can be
used to constrain the ULDM scenario with pulsar timing arrays
\cite{RuKhm, Postnov, newPSRtiming}. In the present work, we explore the
effect of these coherent oscillations on the propagation of
electromagnetic waves through the ULDM condensate. We demonstrate that the
polarization angle of linearly polarized emission oscillates with the same
period, which is determined by the ALP mass and is therefore uniform for
all ALP domains in the Universe. Then, we use long-term observations of
polarization properties of radio sources to search for these oscillations.
We do not find any significant evidence for oscillations with a common
period and use this fact to constrain the photon coupling of the ULDM ALP.

The rest of the paper is organized as follows. In
Section~\ref{sec:calculation}, we consider photon propagation in the
oscillating external ALP background and demonstrate that the plane of
linear polarization of photons oscillates with the same period. In
Section~\ref{sec:data}, we describe the data resulted from long-term radio
observations of parsec-scale jets in active galaxies which we use in this
paper. Section~\ref{sec:anal} presents the method to search, in the
ensemble of data, for oscillations with a common period but arbitrary
phase. In Section~\ref{sec:results}, we present our results and derive
constraints on the ALP-photon coupling. We briefly conclude in
Section~\ref{sec:concl}, while some technical details are presented in
Appendices.

\section{Theoretical calculation of the expected effect}
\label{sec:calculation}
We start with the following Largangian for an ALP interacting with
photons,
\be
\mathcal{L}=
-\frac{1}{4}F_{\mu\nu}^2 +
\frac{1}{2}(\partial_\mu a \partial^\mu a - m^2 a^2) +
\frac{g_{a\gamma }}{4} a F_{\mu\nu}\tilde{F}^{\mu\nu}\,,
\ee
where $a$ is the ALP field, $F_{\mu \nu }$ is the electromagnetic stress
tensor,
$\tilde{F}_{\mu\nu}=\frac{1}{2}\epsilon_{\mu\nu\rho\sigma}F^{\rho\sigma}$
and ALP parameters are the mass $m$ and the photon coupling constant
$g_{a\gamma}$. The latter has the dimension of inverse mass in the natural
($\hbar = c = 1$) system of units, which we hereafter use, unless the
dimensions are written explicitly.
The Minkowski sum over repeating Greek indices $\mu ,\nu ,\lambda ,\rho =
0, \dots, 3$ is assumed (Latin indices $i, j = 1, 2,3$ enumerate
the spatial coordinates).

The equations of motion for the electromagnetic field read
\[
\partial _{\mu } F^{\mu \nu } + \frac{1}{2}
g_{a\gamma } \epsilon^{\mu \nu \lambda
\rho } \partial _{\mu } \left(a F_{\lambda \rho }  \right)=0.
\]
At the scales of order the photon wavelength,
$a$ changes slowly and hence can be treated adiabatically when taking
Fourier integrals. To study the effect of the external $a$ field on the
polarization, we consider a plane-wave Ansatz for the electromagnetic
field,
\[
A_{\nu }(x)=A_{\nu }(k) {\rm e}^{ikx} + \mbox{h.\,c.},
\]
and decompose $A_{\nu }(k)$ into two linearly polarized components,
\[
A_{\nu } =A^{+}e_{\nu }^{+}+A^{-}e_{\nu }^{-},
\]
where $e_{\nu }^{\pm}$ are properly chosen polarization vectors. The
equations of motion result in the dispersion relations (we use here the
fact that $|\partial _{0} a| \sim  m |a|$ while $|\partial _{i}a| \sim mv
|a|$, where $v \ll 1$ is the dark-matter velocity, hence
$|\partial _{0} a| \ll|\partial _{i}a| $),
\[
\omega_{\pm}^{2} -k^{2} \mp  g_{a\gamma } \partial _{0}a |k|=0,
\]
so the two polarization states propagate with
\[
\omega_{\pm} = k \sqrt{1 \pm g_{a\gamma}\frac{\partial _{0}a}{k}}
\simeq
k \pm \frac{1}{2}g_{a\gamma } \partial _{0} a
\]
(see also Refs.~\cite{Dub, Harari:1992ea}).

The axion condensate acts as an optically active medium,
in which a linearly polarized photon acquires a phase
shift between
two circular polarizations, which results in the rotation of the
polarization plane. This effect as a \textit{cosmological
birefringence} has been constrained in the past, see e.g.\
\cite{Wardle:1997gu, Finelli:2008jv, Alighieri:2010eu, Galaverni:2014gca}.
Recent studies~\cite{Obata:2018vvr, DeRocco:2018jwe, Liu:2018icu}
suggested to test this effect with laser interferometry. It should be
noted that these works were considering a constant phase shift
experienced by photons. In contrast, in the present work we focus on the
\textit{periodic} changes of the phase shift caused by the oscillating ALP
background.

The difference between the frequencies of the two polarization components,
\[
\Delta \omega =  g_{a\gamma } \partial _{0}a,
\]
is translated into the change of the polarization angle for a linearly
polarized emission,
\[
\Delta \phi = \frac{1}{2}\int\limits_{t_{1}}^{t_{2}} \! \Delta \omega \, dt =
\frac{1}{2}g_{a \gamma } \int\limits_{t_{1}}^{t_{2}} \!  \partial _{0} a\, dt,
\]
where the integration is performed along the propagation path of the
electromagnetic wave from the emission moment $t_{1}$ to the observation
moment $t_{2}$. We use again
$|\partial _{0} a| \ll|\partial _{i}a| $
to write
\[
\partial _{0} a \equiv \frac{da}{dt} - \frac{k_{i}}{|k|} \partial _{i}a
\simeq \frac{da}{dt}
\]
and therefore obtain
\[
\Delta \phi =\frac{1}{2} g_{a\gamma } \left(a(t_{2})-a(t_{1})   \right).
\]
Note that the effect
 is frequency-independent and depends only on the local value of the ALP
field at the source and the observer. This is true even in the case of
inhomogeneous background~\cite{Harari:1992ea}.

The ALP field is coherent and homogeneous at
the scales of order of
\[
\lambda = \frac{1}{mv}\simeq 65 \left (
\frac{m}{10^{-22}~\text{eV}} \right)^{-1} \left ( \frac{v}{10^{-3}}
\right)^{-1} \,\text{pc}\,,
\]
where $v$ is the mean velocity of dark matter.
It oscillates as
$$
a(t)=a_0\sin(mt+\delta)\,,
$$
where $\delta$ is some random phase and $a_0$ is the field amplitude.
The typical oscillation
period is
\begin{equation}
T=\frac{2\pi}{m}\simeq 4\cdot 10^{7} \left ( \frac{10^{-22}~\text{eV}}{m}
\right)\text{sec}\,.
\label{eq:period}
\end{equation}
Therefore, these background oscillations get imprinted in the oscillations
of the photon phase shift. The observed period at the Earth is
$T'=T(1+z)$, where $z$ is the redshift of the source.

We will be interested in the situation where the ALP field in the
vicinity of the source is much stronger than next to the observer, that
is $a(t_{2}) \ll a(t_{1})$. In particular, this is true for central parts
of elliptical galaxies hosting AGNs, which have a reasonably high
dark-matter density. Indeed, a recent joint analysis of lensing and
kinematics data from Ref.~\cite{Lyskova:2017me} gives the following
estimate for the dark-matter energy density there,
\be
\rho_{\rm DM}\sim 5\cdot
10^9~{M_\odot}/{\text{kpc}^3}\simeq  20~\text{GeV}/\text{cm}^3\,.
\ee
Here
we assumed a typical dark-matter fraction in elliptical galaxies
$\sim\,50\%$ \cite{Lyskova:2014me}. On the other hand, the energy density
\[
\rho_{\rm DM} ={1\over 2} \langle \left( \partial a \right)^{2} \rangle +
\frac{m^{2}}{2} \langle a^{2} \rangle =
{1 \over 2} m^{2} a_{0}^{2},
\]
which we use to express $a_{0}$ through $\rho_{\rm DM}$ and to obtain the
final expression for the oscillating shift of the polarization angle,
\begin{equation}
\Delta \phi \simeq 5^{\circ}
\sin \left(2\pi \frac{t}{T'} +\delta\right)
\left (
\frac{\rho_{\rm DM}}{20~\text{GeV}/\text{cm}^3} \right)^{\frac12} \left (
\frac{g_{a\gamma}}{10^{-12}~\text{GeV}^{-1}} \right) \left (
\frac{m}{10^{-22}~\text{eV}} \right)^{-1}\,.
\label{eq:ampl}
\end{equation}

\section{The data}
\label{sec:data}

\begin{figure}[tbp]
\centering
\includegraphics[width=.85\textwidth,trim=0cm 0cm 0cm
0cm]{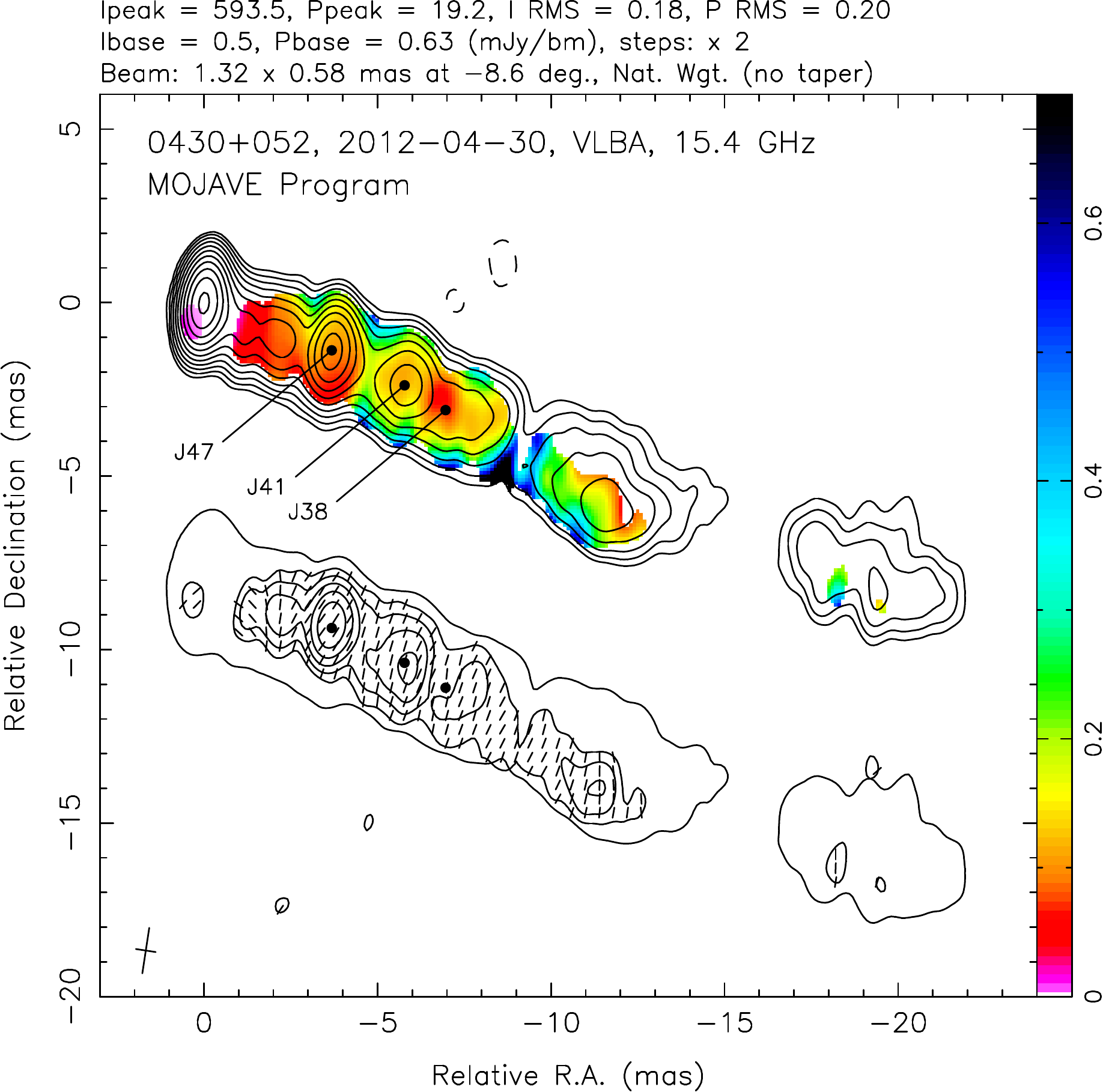}
\caption{\label{fig:3C120}
\blue
Parsec-scale image of the jet in the active galaxy 3C~120 observed by the
Very Long Baseline Array at 2~cm within the MOJAVE program. Top: total
intensity contour plot overlayed by the fractional polarization
distribution represented by the color. Bottom: the lowest contour of the
total intensity distribution showing the shape of the jet overlayed by the
contour plot of the linear polarization map. The sticks represent the
electric vector position angle (EVPA). Black bullets show the positions of
the jet components used in our analysis, with labels according to their
IDs in Table~\ref{tab:sources}. The synthesized VLBA beam is shown at the
half-power level in the bottom left corner.}
\end{figure}

For the purposes of our study, we made use of polarization sensitive
interferometric data at 15~GHz primarily from the MOJAVE (Monitoring Of Jets in
Active galactic nuclei with VLBA Experiments)
program to monitor radio brightness and polarization variations in jets associated with
active galaxies with declinations above $-30^{\circ}$,
\blue with supplementary data obtained from the NRAO archive. \black The
observations were performed \blue between 1997 April 6 and  2017 August 25 \black
with the VLBA (Very Long Baseline Array), a
system of ten 25-meter radio telescopes, allowing to probe
highly-collimated relativistic outflows of the observed sources on parsec
scales by achieving angular resolution of the order of one milliarcsecond.
The fully-calibrated visibility data together with reconstructed FITS
images are publicly available online from the MOJAVE web
site\footnote{\url{http://www.physics.purdue.edu/astro/MOJAVE/allsources.html}}.
\blue
An example of a total intesity and linear polarization image for the
active galaxy 3C~120 is shown in Figure~\ref{fig:3C120}. \black The data
reduction, including initial calibration and editing, was performed with
the NRAO Astronomical Image Processing System \cite{AIPS} using the
standard techniques. Imaging was done with the Caltech DIFMAP package
\cite{difmap}. Each of the final single-epoch images was constructed by
applying natural weighting to the visibility function and a pixel size of
0.1 mas. A more detailed discussion of the data reduction and imaging
process schemes can be found in \cite{MOJAVE_V, MOJAVE_XV}. The earlier
papers of the MOJAVE program have focused on the parsec-scale kinematics
of the jets \cite{MOJAVE_XIII}, their acceleration and collimation
\cite{MOJAVE_XII}, circular and linear polarization properties \blue
\cite{MOJAVE_II, MOJAVE_I, MOJAVE_XV, MOJAVE_XVI, Pushkarev2018}. \black

To analyze the polarization evolution in the observed sources, the
following approach is used. The structure of every source in its full
intensity is modelled using the Caltech DIFMAP package \cite{difmap} by
fitting a series of circular (rarely elliptical) Gaussian components to
the calibrated visibility data. These components are cross-identified at a
certain number of epochs, while they remain sufficiently bright. Their
electric vector position angles, $\mathrm{EVPA} = (1/2)
\mathrm{arctan}(U/Q)$, where $U$ and $Q$ are Stokes parameters, are
calculated from the corresponding polarization maps as the nine-pixel
average of the area centered over a position of the component in total
intensity.
\blue
In this way, jet and core components are identified in Refs.\
\cite{MOJAVE_XIII, Lister-et-al-in-prep}. Because of variable optical
depth
\cite{cs_var}, Faraday depth and higher turbulence \cite{TEMZ}, the core
components demonstrate, on average, larger EVPA fluctuations
\cite{MOJAVE_XVI}. \black In this work, we therefore concentrate only on
the jet features. We estimate that our VLBA EVPA measurements are accurate
within $\sim5^\circ$, as it comes from a comparison of highly-compact AGNs
to near-simultaneous single-dish observations.

The uncertainty in the EVPA measurements is dominated by two factors, see
e.g.\ the Appendix of Ref.~\cite{WardleKronberg}. The first one is the
imperfectness of the receivers which results in the effect of the
instrumental polarization when the signal flows from one branch to
another. The second one is related to the non-uniform coverage of the $uv$
plane. The flow of the polarized signal is proportional to the total
intensity and is therefore non-uniformly distributed over the map. It also
results in an artificial increase of the signal-to-noise ratio, so that
for weak signals one needs to introduce additional corrections for
non-Gaussian distribution of errors. This motivates the removal of
low-intensity sources from the data set to suppress hard to control
instrumental errors. We impose a cut of the minimal polarized flux density
of the component (the sum of polarised emission under the area of a given
Gaussian component, as defined above) of 5~mJy,
which guarantees that the estimated uncertainty in every particular EVPA
measurement is $\sim 5^{\circ}$.

To perform a reliable search for periodicity at the scales of $\sim
1$~yr, as it would be expected for the ULDM effect, we select the source
components for which the 5~mJy condition was satisfied for at least 10
observational epochs and require the cadence of not less than 5 epochs per
year. Among all observations, these criteria are satisfied for one or more
components of 10 sources listed in Table~\ref{tab:sources}, which are
used in this analysis.
\begin{table}[tbp]
\begin{center}
\begin{tabular}{cccccc}
\hline
\hline
B1950 & Other  &  Redshift & Component & Time coverage, & Number of  \\
  name   &   ID   &           &   ID      & years         & epochs \\
\hline
0415$+$379&3C~111&0.0491& \bf 40 &\bf 4.80&\bf 40\\
          &      &      &  39&  3.49  & 19  \\
          &      &      &  30&  1.94  & 19  \\
          &      &      &  35&  1.23  & 13  \\
          &      &      &  36&  1.47  & 19  \\
          &      &      &  37&  2.01  & 13  \\
          &      &      &  39&  3.49  & 19  \\
          &      &      &  57&  1.42  & 11  \\
0430$+$052&3C~120&0.033 &\bf 41&\bf 1.80&\bf 10\\
          &      &      &  38& 1.23   & 11   \\
          &      &      &  47& 0.92   & 11  \\
0851$+$202&OJ~287&0.306 &\bf 22  &\bf 3.15&\bf 16\\
          &      &      &   9&  1.41  &  10 \\
1226$+$023&3C~273&0.1583&\bf 19  &\bf 2.49&\bf 13\\
          &      &      &  24&  1.60  & 13  \\
          &      &      &  25&  1.60  & 12  \\
1253$-$055&3C~279&0.536 &\bf 6 &\bf 2.97&\bf 17\\
1308$+$326&OP~313&0.997 &\bf 5 &\bf 2.57&\bf 13\\
1510$-$089&PKS~1510$-$08&0.36&\bf 15 &\bf 1.65&\bf 14\\
          &      &      &  19& 1.30   & 13  \\
1641$+$399&3C~345&0.593 &\bf 11  &\bf 2.35&\bf 13\\
2200$+$420&BL~Lac&0.0686&\bf 7  &\bf 20.39&\bf 126\\
          &      &      &  20   &   5.09   &  45   \\
          &      &      &  23   &   5.58   &  59   \\
          &      &      &  24   &   2.81   &  25   \\
          &      &      &  25   &   2.55   &  21   \\
          &      &      &  26   &   1.70   &  20   \\
          &      &      &  27   &   1.64   &  25   \\
          &      &      &  36   &   0.66   &  11   \\
          &      &      &  47   &   1.00   &  13   \\
2251$+$158&3C~454.3&0.859&\bf 13  &\bf 1.98&\bf 10\\
          &      &      &  8& 1.92   & 10  \\
\hline
\hline
\end{tabular}
\end{center}
\caption{\label{tab:sources}
List of source components selected for the analysis. MOJAVE ID and
Component ID correspond to the MOJAVE catalog
\blue
\cite{MOJAVE_XIII, Lister-et-al-in-prep}.
\black
Components with the longest time coverage are indicated in bold.}
\end{table}

The separations between jet components in a source are sufficiently large
to make the regions causally disconnected at the period of observations,
therefore making correlated periodic changes of the intrinsic
polarization impossible.
\blue
Other systematic effects could result in common EVPA varations among
components of a given extragalactic radio source. First of all, it is the
EVPA absolute calibration error. This error is the same for all core and
jet components of all sources observed within the same 24-hour long MOJAVE
VLBA epoch. Typically, 20 to 30 targets are observed. The variable
fraction of this error is estimated to be less than $2^\circ$. Second,
polarized emission might partly ``leak'' from one component to another if
they are located too close to each other~--- within one VLBA beam. Such
a situation happens not more than in a couple of cases in our sample.
Finally, let us consider Faraday rotation around an extragalactic jet or
in our Galaxy. Causality arguments prevent synchronous Faraday depth
changes around jets since components are located far enough from each
other, while the Galactic Faraday rotation is low and varies on timescales
significantly longer than what is analyzed in this
paper, see e.g.~\cite{Taylor2009, Opp2012}. We conclude that the common
EVPA variations of components within a given target are insignificant and
cannot influence results of our analysis. We note that the analysis
presented in this paper was repeated for a subsample which has only one
component per target. We obtained qualitatively similar results but with
weaker upper limits, as it is expected from the reduction in statistics.

\section{The analysis method}
\label{sec:anal}
The aim of the present study is to search for, or to constrain, periodic
oscillations of EVPA in different sources but with a common period. The
data analysis we perform includes processing time-dependent
measurements for every particular source in the sample to reveal
indications for periodicity, followed by an analysis of the sample to see
if the periodicities have one common period. We need
a quantitative measure of the strength of the effect, which is
compared to the same quantity obtained from the Monte-Carlo simulated data
assuming no effect is present.

\begin{figure}[tbp]
\centering
\includegraphics[width=.75\textwidth]{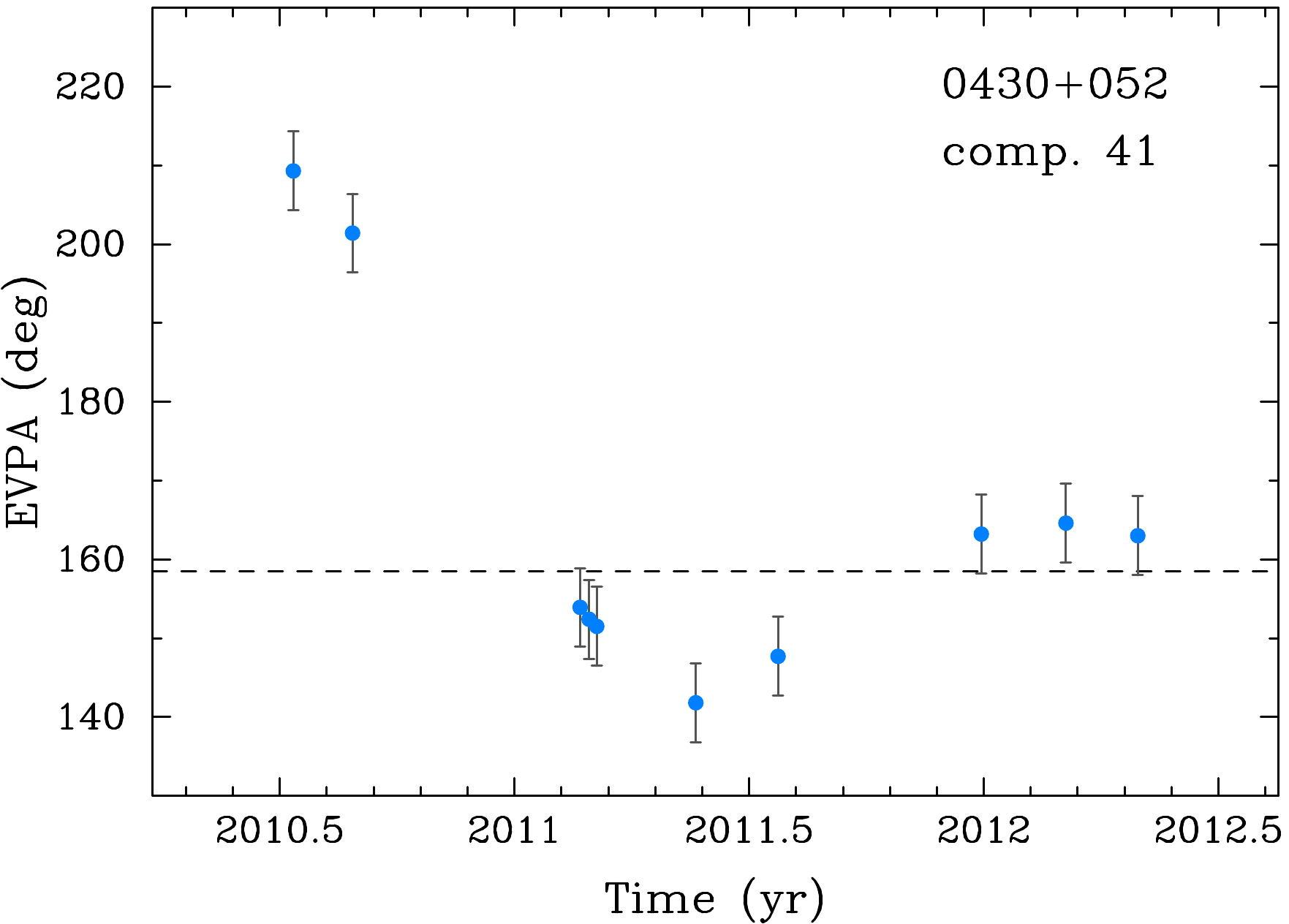}\\
\includegraphics[width=.75\textwidth]{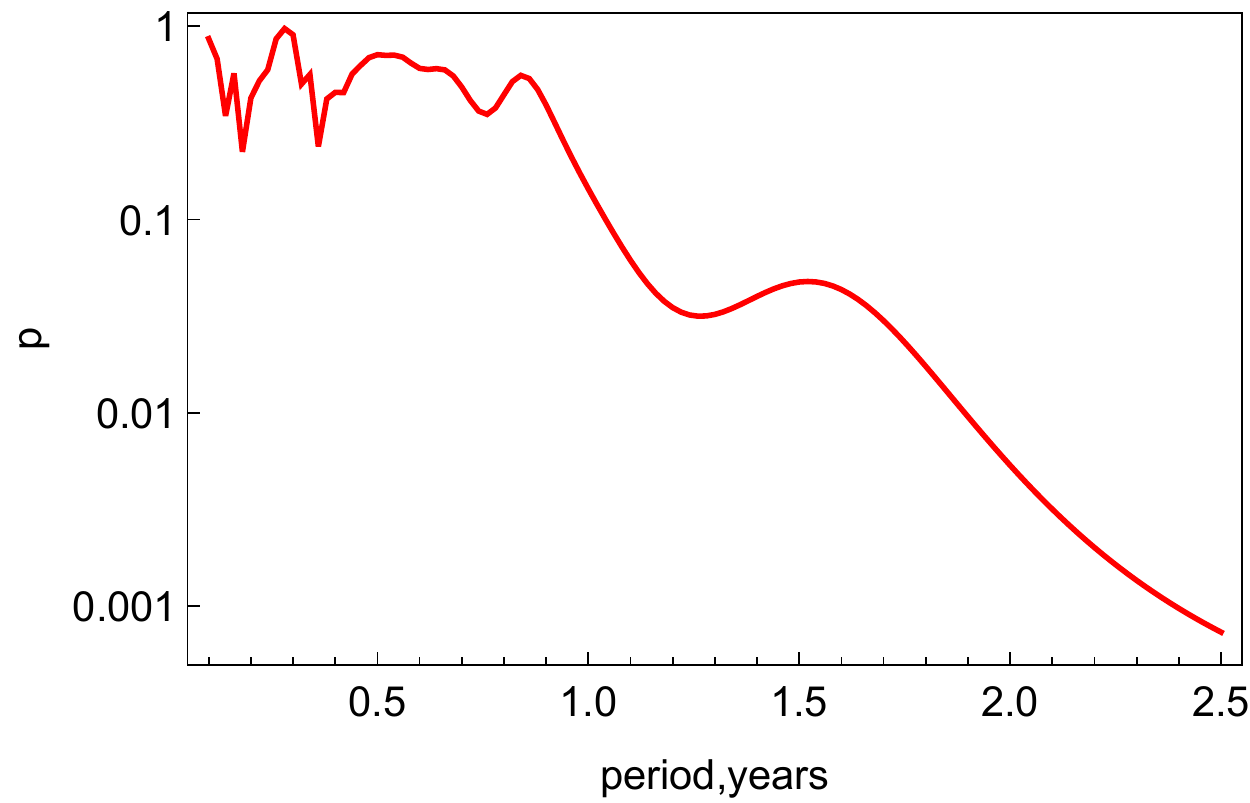}
\caption{\label{fig:1comp}
Example of the data and the periodogram for one
source component. The EVPA observations of component 41 of 3C~120 are
shown versus the observation time (Earth frame, top panel). The $p$-value
calculated from the Lomb--Scargle periodogram for this component is shown
as a function of the period $T$ (source frame, i.e.\ rescaled by $(1+z)$,
bottom panel). Low $p$-values indicate that oscillations with large
periods fit the data well; however, they do not affect the resulting
significance for the full data set estimated through Monte-Carlo
simulations.}
\end{figure}

For every sequence of the EVPA measurements (a source component), we
calculate the generalised Lomb-Scargle periodogram~\cite{periodogr} (see
Refs.~\cite{periodogr17, periodo-book} for detailed discussions). Compared
to other methods, it is more suitable for the case when measurements are
distributed non-uniformly in time. It also accounts correctly for a
non-zero mean of the measurements. For convenience, the calculation of the
periodogram is summarized in Appendix~\ref{app:periodogram}. As a result,
we obtain $p$-values as a function of the assumed oscillation period $T$
(rescaled from the observed $T'$ by $(1+z)$, see
Sec.\ref{sec:calculation}). We consider periods between 0.1~yr and 1.5~yr
with a step of 0.02~yr. This corresponds to the ULDM particle mass $5
\times 10^{-23}$~eV$\le m \le 1.2 \times 10^{-21}$~eV, see
Section~\ref{sec:calculation}. The meaning of the $p$-value is the
probability
that a signal of a given power (or stronger) is produced by random
Gaussian fluctuations;
note that $0 \le p \le 1$. In practice, relevant random backgrounds are
not distributed normally, and therefore we do not interpret this quantity
as a probability. Figure~\ref{fig:1comp}
presents an example of the data and the periodogram for one of the source
components shown in Figure~\ref{fig:3C120}.

Next, we need to consider the ensemble of $N$ time sequences ($N=32$ in
our case). For each time sequence, i.e.\ for each source component, we have
the dependence of the $p$-value on the period, $p_{i}(T)$, $i=1,\dots, N$.
For every particular source, small $p_{i}(T)$ indicates that the data
favour, to some extent, oscillations with a period $T$. Consider now the
function
$$
L(T)=\log \prod\limits_{i=1}^N p_i (T)=
\sum\limits_{i=1}^N \log \left(p_i (T)  \right).
$$
Qualitatively, low $L(T)$ would indicate  periodic
signals with \textit{the same period} $T$ in different sources, as it is
expected in ULDM models. Contrary, if periodicities in individual sources
are absent or  uncorrelated, as it is expected for their intrinsic origin,
minima of $p_{i}(T)$ would not coincide, and $L(T)$ would not be that low.

What does a particular value of $L(T)$ mean? To understand it in terms of
physical quantities, we perform Monte-Carlo simulations of many artificial
data sets, both without the ULDM effect and with the EVPA oscillations with
the common period and amplitude introduced by hand, for various periods
and amplitudes. The simulations are described in detail in
Appendix~\ref{app:MC}. The simulations of random sets,
Section~\ref{app:MC:MC}, allow one to answer the question how often a
given value of $L(T)$ may be obtained as a random fluctuation of the data
with no signal for a given $T$, thus attributing a local $p$-value to the
observed realization of $L(T)$ for the ensemble of sources. The global
significance of the observed deviation from randomness is estimated
through the same simulation as a measure of the fraction of MC sets for
which this or lower $p$-value is obtained from fluctuations for
\textit{any} $T$.

To convert $L(T)$ into a limit on the amplitude of EVPA oscillations
$\phi$, we use the simulations described in Section~\ref{app:MC:expected},
when signals with various $\phi$ are artificially added to the random data.
From these simulations we determine for which $\phi$ a given or lower
value of $L(T)$ occurs in 95\% of simulated data sets. This allows us to
derive the 95\% CL upper limit on $\phi$ for every $T$. These
upper limits on the common amplitude $\phi$ at a given common period $T$
are finally translated into the limits on the axion-photon coupling
$g_{a\gamma}$ at a given ALP mass $m$ using Eqs.~(\ref{eq:period}),
(\ref{eq:ampl}) derived in Section~\ref{sec:calculation}.

\section{Results and discussion}
\label{sec:results}
\begin{figure}[tbp]
\centering
\includegraphics[width=.85\textwidth]{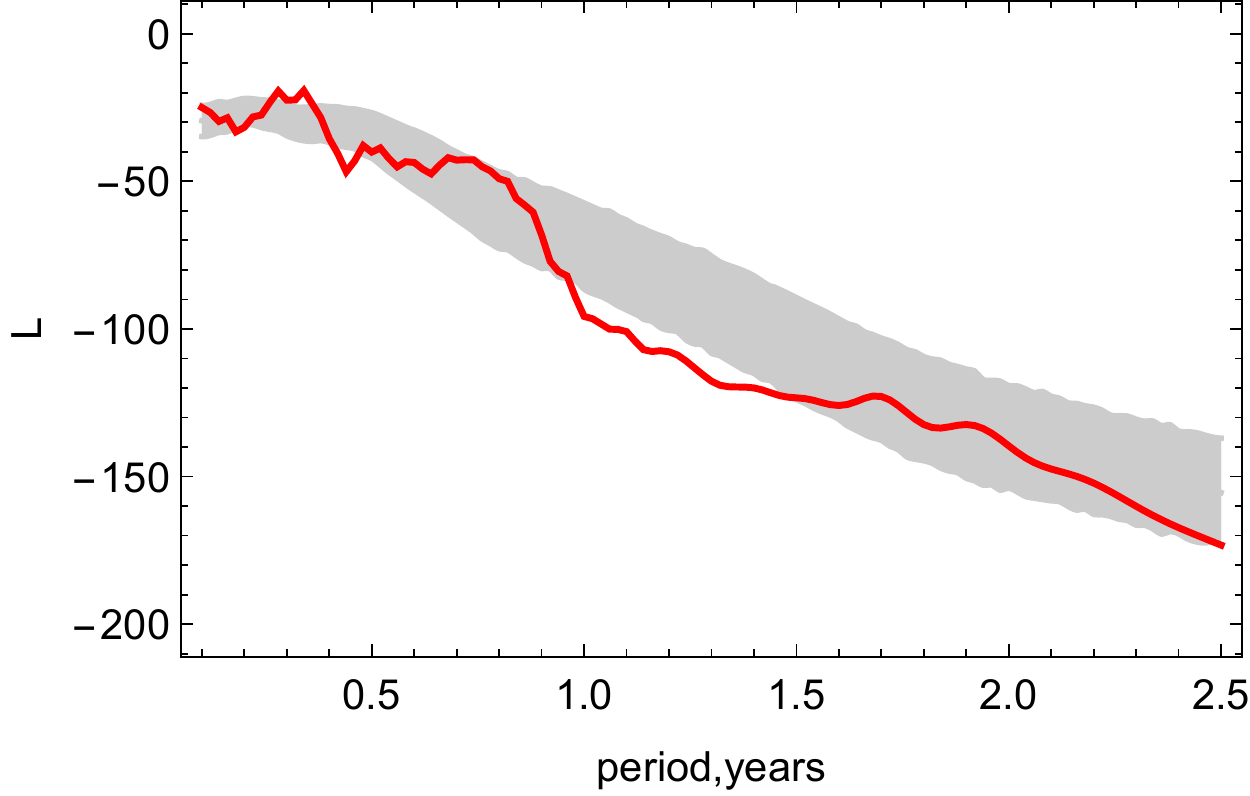}
\caption{\label{fig:result1}
Expected \blue in the case of no ULDM effect \black
(gray band, 68\% CL) and observed (thick red line) $L$ as a
function of the period $T$.
}
\end{figure}

\begin{figure}[tbp]
\centering
\includegraphics[width=.75\textwidth]{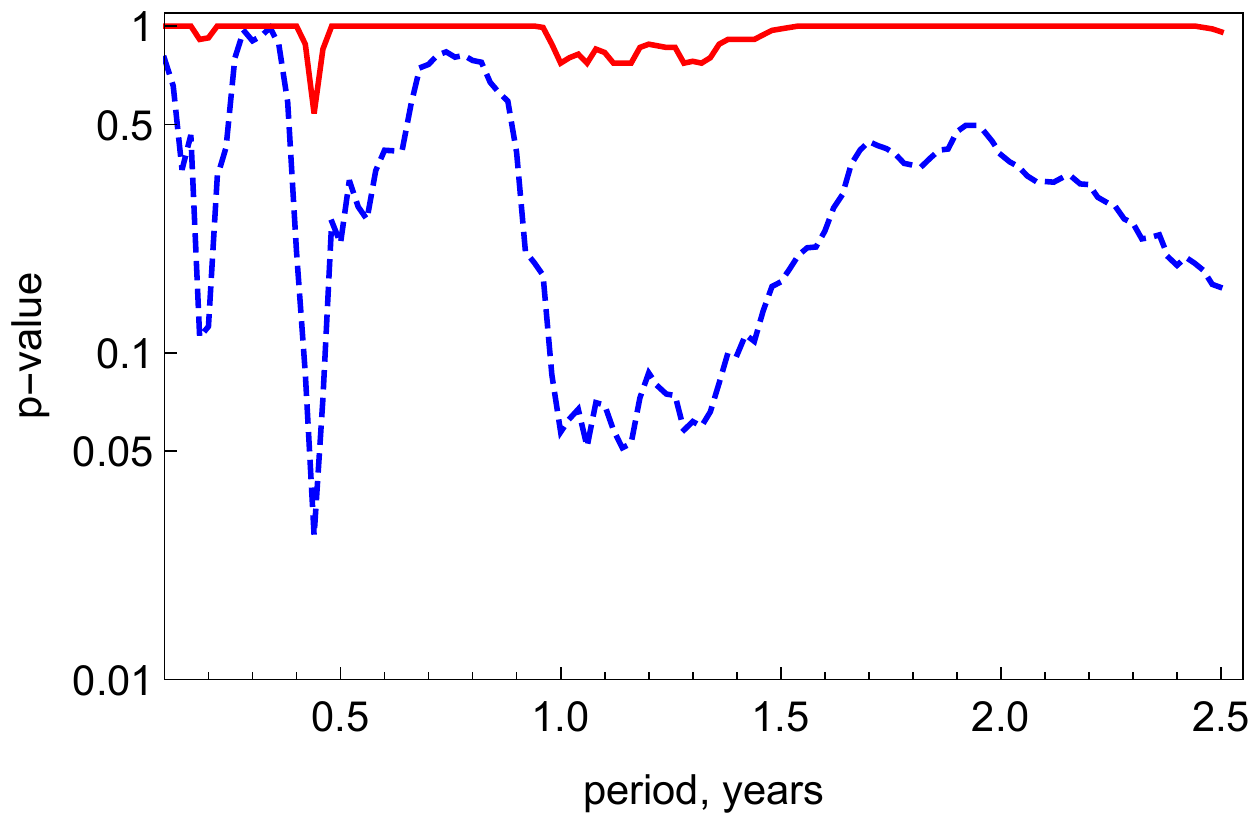}
\caption{\label{fig:pval}
Local (dashed blue line) and global (full red line) p-values for deviations
of the observed value of $L$ from the expected one, as a function of the
period $T$.}
\end{figure}

\begin{figure}[tbp]
\centering
\includegraphics[width=.85\textwidth]{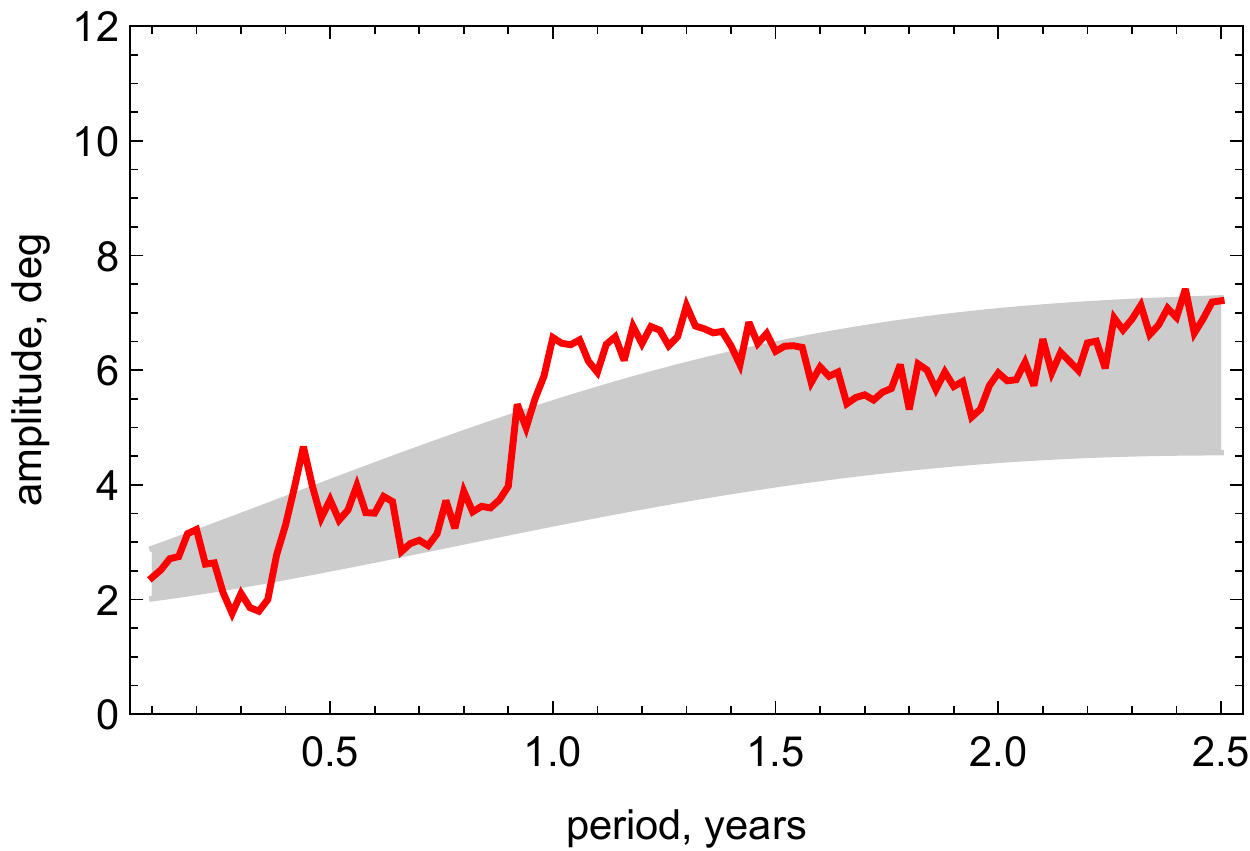}
\caption{\label{fig:result2}
Expected in the case of no ULDM effect (gray band, 68\% CL) and
observed (thick red line) 95\% CL upper
limits on the amplitude of oscillations $\phi$ as a function of the period
$T$. }
\end{figure}

\begin{figure}[tbp]
\centering
\includegraphics[width=.85\textwidth]{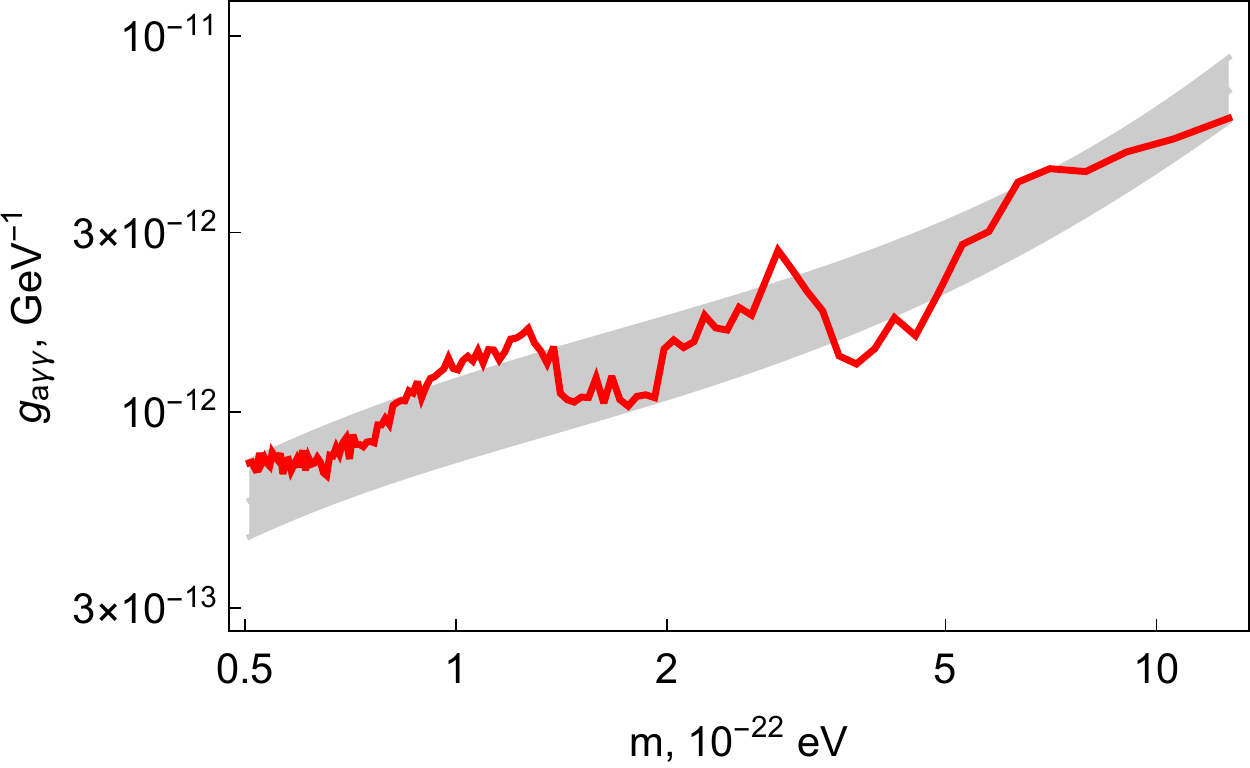}
\caption{\label{fig:result3}
Expected in the case of no ULDM effect (gray band, 68\% CL) and observed
(thick red line) 95\% CL upper
limits on the axion-photon coupling $g_{a\gamma}$ as a function of the
ALP mass $m$, for $\kappa=1$.
}
\end{figure}

We turn now to the results of the data analysis. They are presented
in Figures~\ref{fig:result1}, \ref{fig:pval}, \ref{fig:result2},
\ref{fig:result3}.

Figure~\ref{fig:result1} presents
values of $L$ for various
oscillation periods $T$, as
expected from MC simulations (for the
case of no ULDM effect) and observed in real data.
The observed $L(T)$ is in good
agreement with expectations, indicating that no ULDM effect is seen.
Figure~\ref{fig:pval} confirms this by presenting $p$-values which
correspond to probabilities of obtaining the observed or lower $L(T)$ in
Monte-Carlo simulated sets: the local $p$-value corresponds to a given $T$
while the global $p$-value is the probability to observe this or lower
local $p$-value for any $T$. The corresponding 95\% CL upper limits on the
amplitude $\phi(T)$ are given in Figure~\ref{fig:result2}. They are
interpreted in terms of physical parameters of ALP in
Figure~\ref{fig:result3}. Note that the conversion of the limits on the
EVPA oscillation amplitude versus $T$ into the limits on $g_{a\gamma}(m)$,
that is the transition from Figure~\ref{fig:result2} to
Figure~\ref{fig:result3}, is subject to theoretical systematic
uncertainties, the main of which is the lack of knowledge of the
dark-matter density in particular observed sources, encoded in the
parameter $\kappa \equiv \rho_{\rm DM}/(20~\rm GeV/cm^{3})$. Our limits on
$g_{a\gamma }$ scale with $\kappa^{1/2}$.

We turn now to the comparison of our results with other available
constraints on $g_{a\gamma}$ for ULDM ALPs.
One should note here that the scalar ULDM with masses below $\sim
10^{-21}$~eV is disfavoured by the Lyman-alpha forest measurements
\cite{1703.04683}. However, these constraints are not applicable for
certain models with ULDM ALPs, see e.g.\ Ref.~\cite{1810.05930}.

\begin{figure}[tbp]
\centering
\includegraphics[width=.65\textwidth]{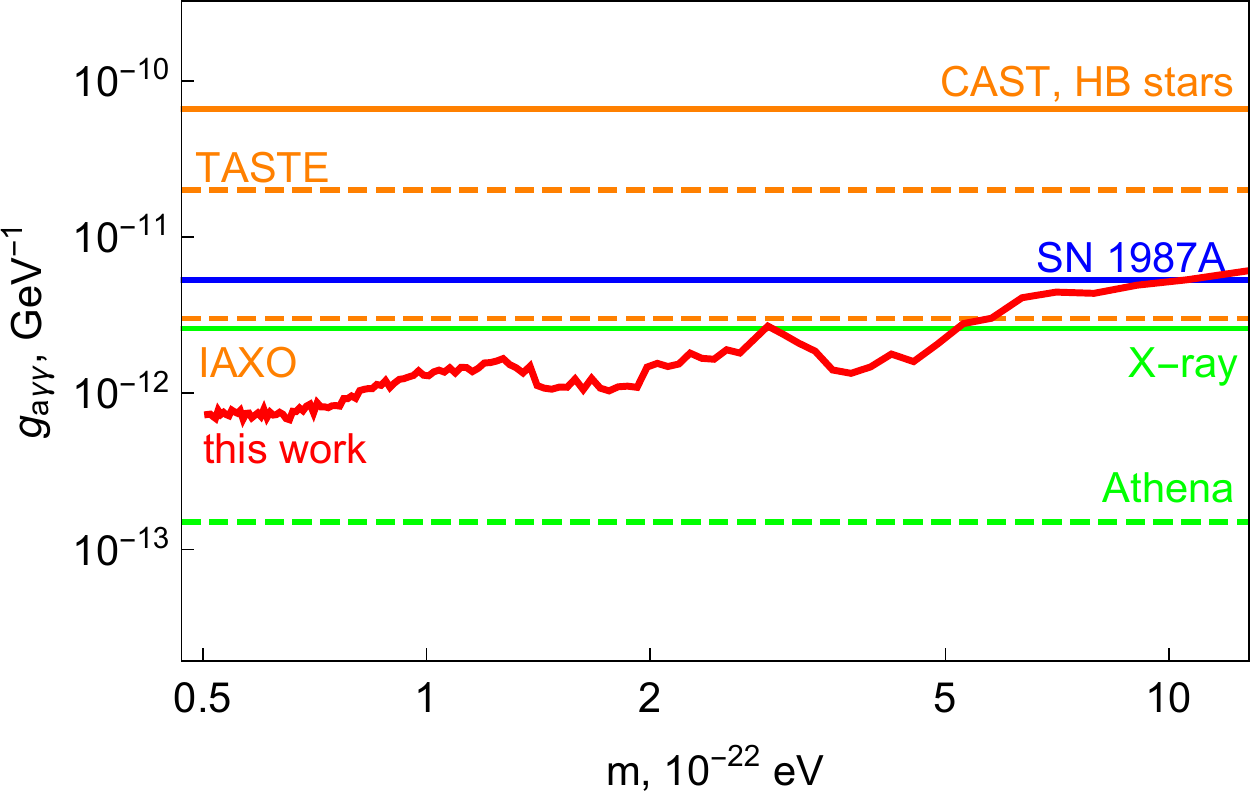}
\caption{\label{fig:limits2}
Upper
limits on the axion-photon coupling $g_{a\gamma}$ as a function of the
ALP mass $m$ from astrophysical effects (not necessary assuming ULDM):
solar axion searches with CAST \cite{1705.02290}, energy losses of the
horizontal-branch (HB) stars \cite{1406.6053} (quantitatively the
same as CAST), absence of gamma rays from SN~1987A \cite{1410.3747} and
absence of spectral irregularities in the X-ray spectrum of Vir~A
\cite{1703.07354} (see the text for more references and discussions of
caveats). Projected sensitivities of axion helioscopes TASTE
\cite{1706.09378} and IAXO \cite{1103.5334}, as well as of the search for
X-ray spectral irregularities with Athena \cite{1707.00176} are shown by
dashed lines. The limit from the present work  ($\kappa=1$) is given by the
full red line for comparison.}
\end{figure}

The most abundant group of constraints includes those based on
astrophysical effects of ALPs independent on whether they form the dark
matter or not. The least model dependent bound results from
non-observation of ALPs coming from the Sun with a
dedicated axion helioscope, CAST~\cite{1705.02290}. A quantitatively
similar constraint comes from the analysis of energy losses in
horizontal-branch stars in globular clusters~\cite{1406.6053}. A stronger,
but more model dependent constraint was derived from non-observation of
gamma rays from supernova SN~1987A~\cite{1410.3747}. A series of
constraints have been obtained from the absence of spectral irregularities
of X-ray sources embedded in the galaxy clusters, see
Refs.~\cite{1304.0989, 1605.01043, 1703.07354, 1704.05256}. These
constraints are heavily based on the modelling of the magnetic fields in
the clusters which is far from being certain. One of the strongest
constraints comes from the nearby Virgo cluster~\cite{1703.07354} for
which the turbulent component of the magnetic field was modelled with a
certain level of confidence\footnote{A slightly stronger constraint was
reported in Ref.~\cite{1704.05256} for a Seyfert galaxy 2E~3140 in the
Abell~1795 cluster; however, the precise location of this galaxy within
the cluster is unknown.}. In all these studies, a possible regular
component of the cluster magnetic field was ignored, which makes the
conclusions less robust. The astrophysical constraints are compared to our
results in Figure~\ref{fig:limits2},
where we also show projected sensitivities of helioscope experiments TASTE
\cite{1706.09378} and IAXO \cite{1103.5334}, as well as the expected
sensitivity of X-ray irregularity analysis with the future instrument
Athena \cite{1707.00176}.

\begin{figure}[tbp]
\centering
\includegraphics[width=.65\textwidth]{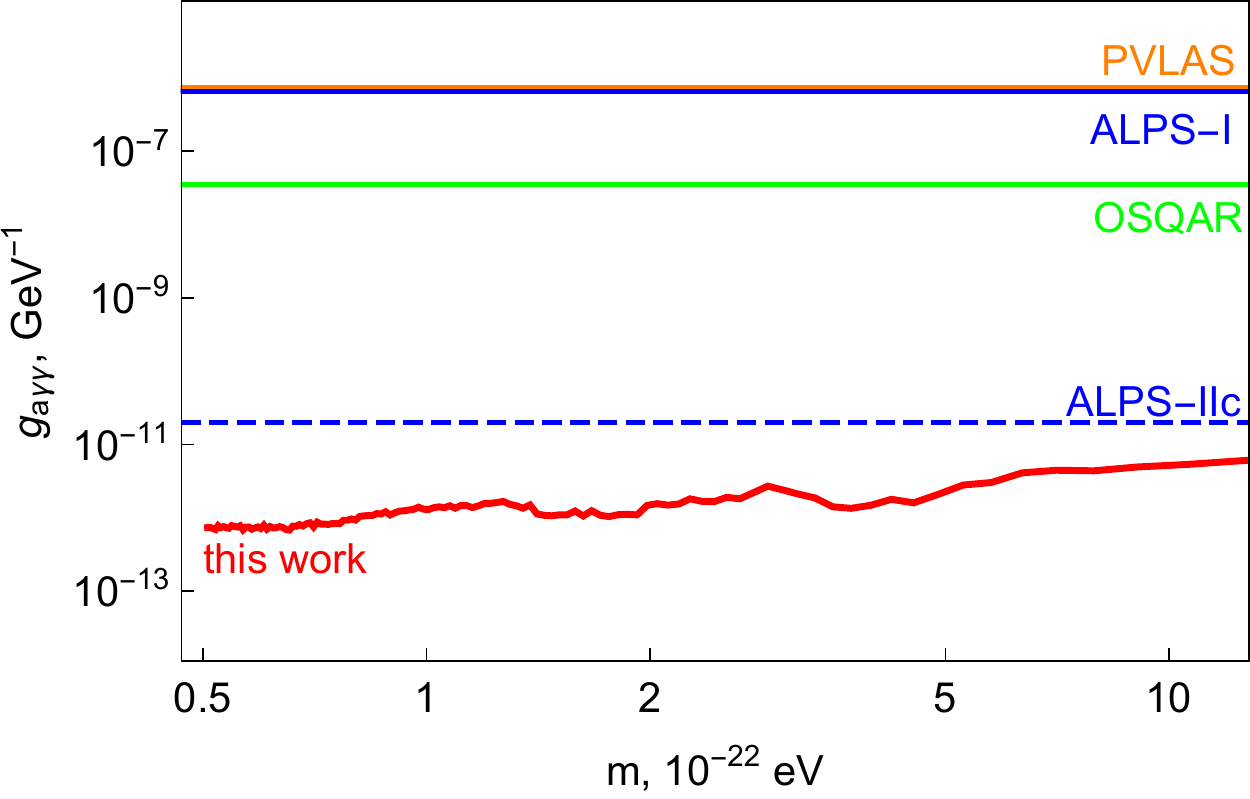}
\caption{\label{fig:limits3}
Upper
limits on the axion-photon coupling $g_{a\gamma}$ as a function of the
ALP mass $m$ from purely laboratory searches: ALPS-I \cite{1004.1313},
PVLAS \cite{1510.08052} and OSQAR \cite{1506.08082}.
Projected sensitivity of ALPS-IIc \cite{1302.5647} is shown by the dashed
line. The limit from the present work ($\kappa=1$) is given by the full red
line for comparison.}
\end{figure}

Perfectly model-independent results come from purely laboratory
experiments, which include ``light shining through walls'' (e.g.\ ALPS-I
\cite{1004.1313} and OSQAR~\cite{1506.08082}) and searches for vacuum
birefringence, PVLAS~\cite{1510.08052}. Though this approach is the most
robust, it results in constraints on ALPs several orders of magnitude
worse than astrophysical ones, just because of limitations of the
terrestrial equipment as compared to astrophysical environments. The
laboratory constraints, as well as the ones expected from the
resonant-regeneration ALPS-IIc experiment~\cite{1302.5647}, are compared
to our results in Figure~\ref{fig:limits3}.

\section{Conclusions}
\label{sec:concl}
Certain dark-matter models assume that ultra-light axion-like particles
form a condensate, coherently oscillating with a period of about a year in
domains of about 100~pc. These oscillations get imprinted on the
polarization angle of linearly polarized electromagnetic emission of
astrophysical sources. The period of the oscillations in the source frame
is universal and is determined by the particle mass only, so that the
corresponding periodic changes would be present for various sources with
the same period.

In the present work, we used the data obtained in long-term
polarization measurements of parsec-scale jets in active galactic nuclei
within the MOJAVE project to search for such periodic patterns with a
common period. We did not find any statistically significant effect and
have obtained constraints on the photon coupling to axion-like ultralight
dark matter
at the level of
$\lesssim 10^{-12}$~GeV$^{-1}$ for masses between $\sim 5\times
10^{-23}$~eV and $\sim 1.2 \times 10^{-21}$~eV.


\acknowledgments
We are indebted to Grigory Rubtsov, Sergey Sibiryakov, Guenter Sigl and
Peter Tinyakov for interesting and helpful discussions and to Eduardo Ros
for useful comments on the manuscript. This research has made use of data
from the MOJAVE data base, which is maintained by the MOJAVE team
\cite{MOJAVE_XV}. The MOJAVE project was supported by NASA-\textit{Fermi}
GI grants NNX08AV67G, NNX12A087G, and NNX15AU76G. Y.K.\ and A.B.P.\ are
partly supported by the Russian Foundation for Basic Research (project
17-02-00197) and the government of the Russian Federation (agreement
05.Y09.21.0018). \blue T.S. was funded by the Academy of Finland projects
274477, 284495 and 312496. \black The work of M.I.\ and S.T.\ on
constraining parameters of ultra-light axions with astrophysical methods
is supported by the Russian Science Foundation (grant 18-12-00258). S.T.\
thanks CERN Theory Department for hospitality at the final stages of this
work.

\paragraph{Note added.}
When this study was finalized, two preprints have been posted on arXiv
which study constraints on $g_{a\gamma }$ for ULDM ALPs within
different approaches. Like our study, they are based on polarization
effects, but do not consider the periodic oscillations which are at the
base of our method. They include the search for birefringence in
observations of the protoplanetary disk of AB~Aur \cite{1811.03525} and
the Cosmic Microwave Background \cite{1811.07873} at the
cosmological and Galactic scales. They report upper limits of $g_{a\gamma
} \lesssim 10^{-13}$~GeV$^{-1}$ for $m\sim 10^{-22}$~eV. While a detailed
discussion of these results is beyond the scope of the present paper, we
note that both Refs.~\cite{1811.03525} and \cite{1811.07873} assume an
additional enhancement of the polarization-angle rotation by a factor of
$\sqrt{n}$, where $n$ is the number of $\sim 100$~pc ALP-field domains
crossed by the light on its way from the source to the observer. This
enhancement is at odds with earlier works~\cite{Dub, Harari:1992ea} (in
more detail, corrections to the results of Ref.~\cite{Harari:1992ea} will
be discussed elsewhere \cite{IvanovPanin}). This $\sqrt{n}$ factor is
close to one for AB~Aur, which is only 163~pc away; however, the
observations \cite{obs:proto} used in Ref.~\cite{1811.03525} continued
only for $\sim 3$~min and therefore give only a snapshot of possible EVPA
oscillations. Observations of this kind would be a very prospective tool
to constrain $g_{a\gamma}$ if they were performed at several epochs and
for several sources, like those used in the present paper.

\appendix
\section{Generalized Lomb--Scargle periodogram}
\label{app:periodogram}
In this Appendix, we collect, for convenience, the formulae used to derive
the generalized Lomb--Scargle periodogram, following
Ref.~\cite{periodogr}.

Consider a series of $N$ measurements of a quantity $y_{i}
\pm \sigma_{i}$, performed at epochs $t_{i}$, $i=1,\dots N$. Introduce
vectors of the time arguments $\bm{t}=\{t_{i}\}$, data values
$\bm{y}=\{y_{i}\}$ and inverse errors $\bm{\bar\sigma}=\{1/\sigma_{i}\}$.
Determine the vector of normalized weights $\bm{w}=\{w_{i}\}$ with
$$
w_i=\frac{1}{\bm{\bar\sigma} \cdot \bm{\bar\sigma}}\,\frac{1}{\sigma_i^2},
$$
satisfying $\sum w_{i}=1$ (hereafter the vectors are denoted by bold face,
the dot between two vectors denotes their scalar product while the dot
between two scalars denotes their product).

For every period $T$ we want to consider, denote $\bm{C}= \{\cos(\omega
t_{i}) \}$ and $\bm{S}= \{\sin(\omega t_{i}) \}$, where $\omega=2\pi/T$.
One further denotes:
$$
Y=\bm{w} \cdot \bm{y}, ~~ C=\bm{w} \cdot \bm{C}, ~~ S=\bm{w} \cdot \bm{S},
$$
$$
\hat Y_Y=\sum w_i y_i^2,
~~
Y_Y=\hat Y_Y- Y \cdot Y,
$$
$$
\hat Y_C=\sum w_i y_i c_i,
~~
Y_C=\hat Y_C- Y \cdot C,
$$
$$
\hat Y_S=\sum w_i y_i s_i,
~~
Y_S=\hat Y_S- Y \cdot S,
$$
$$
\hat C_C=\sum w_i c_i^2,
~~
C_C=\hat C_C- C \cdot C,
$$
$$
\hat S_S=1-\hat C_C,
~~
S_S=\hat S_S- S \cdot S,
$$
$$
\hat C_S=\sum w_i c_i s_i,
~~
C_S=\hat C_S- C \cdot S,
$$
$$
D= CC \cdot SS - CS \cdot CS.
$$
The power spectrum (``normalized periodogram'') is then determined as
$$
P(\omega) = \frac{1}{YY \cdot D} \left( SS\cdot YC \cdot YC + CC \cdot YS
\cdot YS -2\, CS \cdot YC \cdot YS \right).
$$
One notes that $0 \le P(\omega) \le 1$. The local $p$-value for a given
frequency $\omega$ is
$$p(\omega)=\left(1-P(\omega) \right)^{\frac{N-3}{2}}.$$
To obtain the global significance of a minimum in $p(\omega)$, one needs a
more complicated account of the look-elsewhere correction, which may be
achieved, for instance, by the Monte-Carlo simulations.

\section{Monte-Carlo simulations and expected limits}
\label{app:MC}
\subsection{Monte-Carlo simulation}
\label{app:MC:MC}
To perform statistical studies of the data used in this work, we need to
compare actual results with those expected for random data sets, in
generation of which one assumes the absence of the effect we are looking
for. Generation of these sets is not straightforward because the actual
data may have intrinsic non-randomness not related to the effect of ULDM.
In particular, intrinsic conditions in the sources may induce periodic or
quasi-periodic oscillations of EVPA, see e.g.\
Refs.~\cite{period1, period2}: we do not know in advance whether these
features are present or not. Therefore, assuming fully random values of
EVPA based on the experimental error bars would be misleading: to imitate
the ULDM effect in such MC sets, (i)~periodic fluctuations should appear
in various sources and (ii)~their periods, by chance, should coincide. If,
however, intrinsic periodic background is present in particular sources,
the true probability to initiate the signal from fluctuations is
determined by (ii) only. We therefore need to keep unknown features,
including periodic ones, in the simulated data sets.

To this end, we adopt a time-scaling procedure: to generate MC time series
of EVPA measurements for a source, we take the actual data $(t_{i},y_{i})$
and introduce the factor $\xi$, a random number uniformly
distributed between 0.5 and 2.5, by which the observational time is
scaled. The simulated data set is then $(t'_{i},y_{i})$, where
$$
t'_i=t_1 +\xi (t_i - t_1)
$$
(the measured values of EVPA, $y_{i}$, are taken from the actual data
set). The value of $\xi$ is chosen randomly for every source in the
sample. In this way, any potential ULDM signal (periodic changes with a
common period for all sources in the set), which might be present in the
data, is removed while any intrinsic features, including individual
periodicity, remain.

\subsection{Estimate of the expected limits}
\label{app:MC:expected}
\begin{figure}[tbp]
\centering
\includegraphics[width=.65\textwidth]{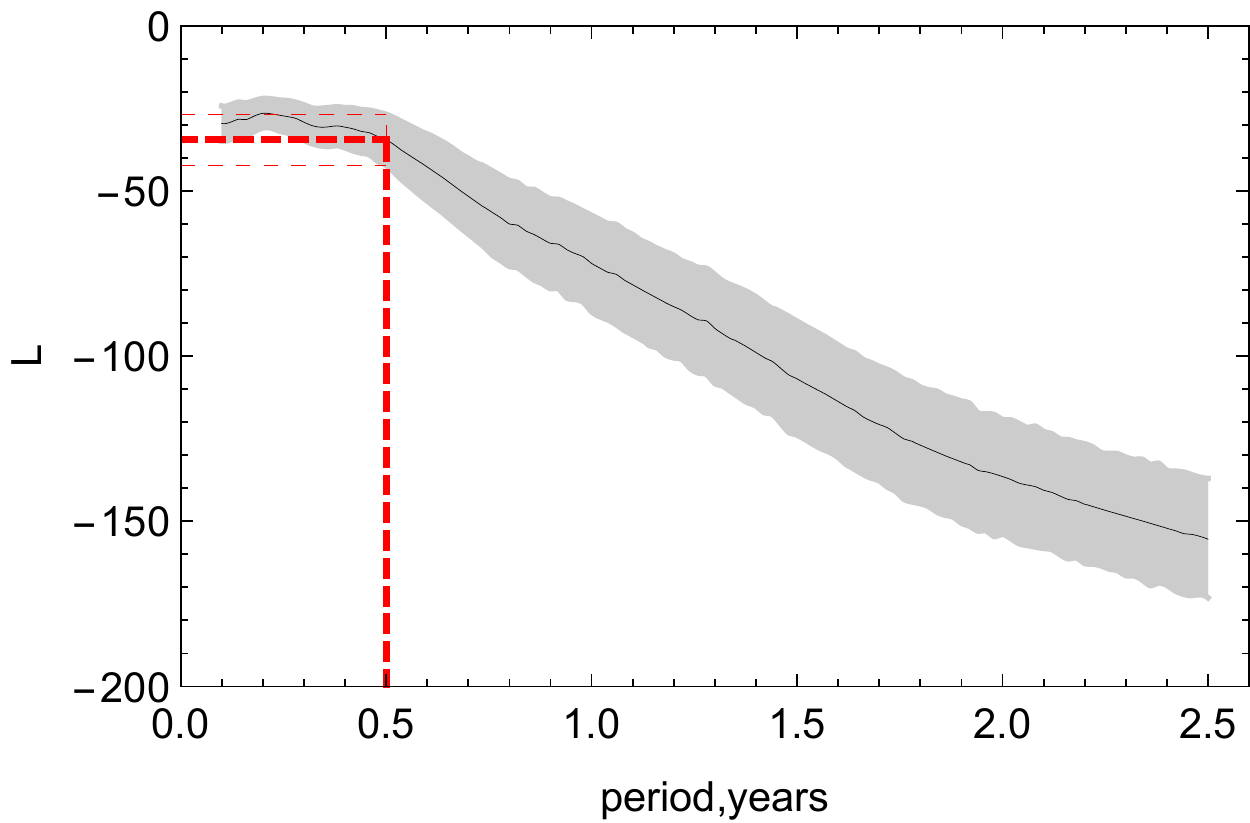}\\[5mm]
\includegraphics[width=.65\textwidth]{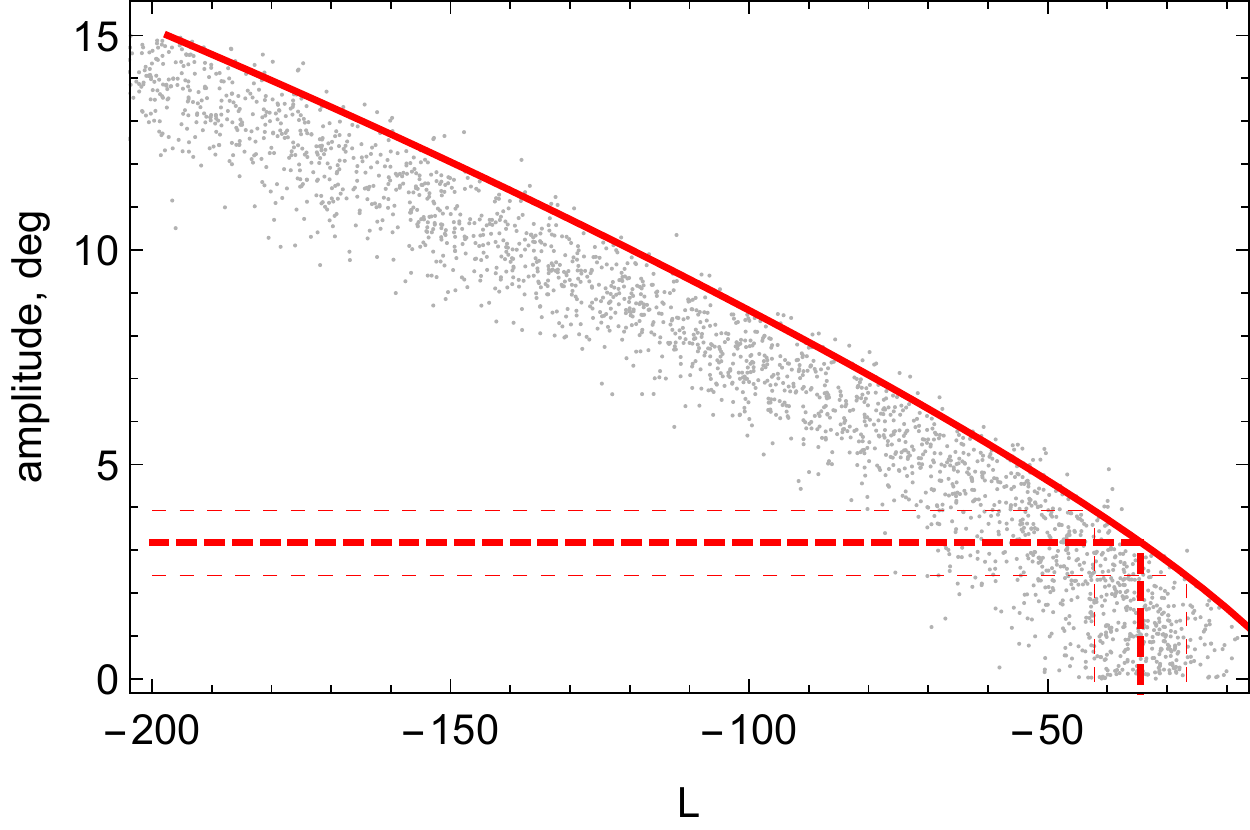}
\caption{\label{fig:expected}
Calculation of the expected limit based on MC simulations. The upper panel
presents the scatter of $L(T)$ in random MC samples (the full line is
the mean value and the gray band contains 68\% of MC points for a given
$T$). For example, for $T=0.5$~yr, the mean value and the 68\% band are
shown by thick and thin dashed red lines, respectively. The lower panel
is based on the Monte-Carlo samples with artificially introduced periodic
oscillations of EVPA with a period of 0.5~years and various amplitudes
and phases. For a given value of $L$, 95\% of the points are below the
full red line. The determination of the mean expected 95\% CL upper limit
and its 68\% CL expected range, based on the values of $L$ obtained from
the upper panel, is shown by dashed lines.}
\end{figure}

To estimate the limits on the amplitude $\phi$ of periodic oscillations of
the polarization angle, we use the following procedure. We generate a
large number (2500 for each value of the period $T$ considered) of MC data
sets assuming no common periodic effects are present beyond those
generated randomly. For every artificial data set, we reconstruct and
record the value of $L$, as described in Sec.~\ref{sec:anal}, as a
function of $T$. For each given $T$, we determine the 68\% CL band,
$\Delta L_{68}(T)$, and the mean, $L_{\rm mean}(T)$, of the values of $L$
which assume no signal (see Figure~\ref{fig:expected}, upper panel).
This is the band
presented in Figure~\ref{fig:result1}
of the main text together with the function $L(T)$ obtained for the real
data.

Next, we perform another MC simulation, this time assuming that some ULDM
effect is present. For each fixed period $T$, we generate 2500 MC data
sets as described above, but with artificially added harmonic oscillations
with period $T$, amplitude $\phi$ (fixed for all objects in the set but
random from one MC set to another, following a uniform distribution
between $0^\circ$ and $15^{\circ}$) and phases $\delta_{i}$ (random for
every object in every MC set). Then, still for the fixed $T$, we determine
the 95\% upper limit on $\phi$ as a function of $L$ as follows. Take the
interval $(L-\epsilon, L+\epsilon)$ for a sufficiently small $\epsilon$
($\epsilon=5$ was used) to have enough data points in this interval and
find the value $\phi_{95}$ such that 95\% of the MC points in the interval
have $\phi<\phi_{95}$. This, for each given $T$, allows us to obtain the
function $\phi_{95}(L)$ which we fit by a smooth 4-parametric curve to
suppress fluctuations related to particular MC realizations. This function
is presented in Figure~\ref{fig:expected} (lower panel) as a thick red
curve for a particular value of $T=0.5$~yr. The values of
$\phi_{95}(L_{\rm mean})$ and $\phi_{95}(\Delta L_{68})$, with the values
of the arguments obtained at the previous step, give the mean value and
the 68\% CL band for the expected 95\% CL upper limit on $\phi$ for a given
$T$, assuming that no ULDM effect is present. This results in the band of
expected upper limits as a function of $T$, presented in
Figure~\ref{fig:result2} of the main text together with the limits
obtained from the real data in the same way.

\end{document}